# ORIENTATIONAL AND STERIC EFFECTS IN LINEAR ALKANOATES + *N*-ALKANE MIXTURES

Juan Antonio González,* Fernando Hevia, Luis Felipe Sanz, Daniel Lozano-Martín, Isaías. García de la Fuente, José Carlos Cobos

G.E.T.E.F., Departamento de Física Aplicada, Facultad de Ciencias, Universidad de Valladolid, Paseo de Belén, 7, 47011 Valladolid, Spain.

*corresponding author, e-mail: jagl@termo.uva.es; Fax: +34-983-423136; Tel: +34-983-423757

**ABSTRACT**

The $CH_3(CH_2)_uCOO(CH_2)_vCH_3$ + *n*-alkane mixtures have been investigated on the basis of an experimental database containing effective dipole moments of esters, and excess molar functions of the systems: enthalpies ($H_m^E$), volumes ($V_m^E$), isobaric heat capacities ($C_{pm}^E$) and isochoric internal energies ($U_{Vm}^E$) and by means of the application of the Flory model and the Kirkwood-Buff formalism. The situation of the mixtures within the $G_m^E$ (excess molar Gibbs energy) vs. $H_m^E$ diagram has also been briefly considered. Results indicate that dispersive interactions are dominant and that steric effects can explain some differences between solutions containing heptane and isomeric esters. Proximity and orientational effects are also discussed in diester + hexane mixtures. In the case of systems with a given alkane and different isomeric polar compounds, orientational effects become weaker in the order: *n*-alkanone > dialkyl carbonate > *n*-alkanoate. Results from the Kirkwood-Buff formalism indicate that the number of ester-ester interactions decreases in systems with alkyl ethanoates when the alkyl size increases and that preferential solvation between polar molecules decreases as follows: dialkyl carbonate > *n*-alkanone > *n*-alkanoate.

## 1. Introduction

Esters are interesting compounds due to their many applications. It is well known that they are used in food and cosmetics industries where their pleasant odors and antifungal properties are important [1,2]. Bio-based polyesters, such as polyhydroxy alkanoates, are an alternative for petrochemical plastics since they can be synthesized by microbial cultures grown on renewable materials in clean environments [3-5]. On the other hand, monoalkyl esters of long chain fatty acids have gained interest as liquid biofuels in view of their great potential of $CO_2$ and $SO_2$ emission reductions [6-8].

Different versions of UNIFAC [9,10], DISQUAC [11,12] or the Nitta-Chao [13] models have been used for the theoretical characterization of *n*-alkanoate + alkane mixtures [14-19]. However, no systematic study is available in the literature about interactions in the mentioned solutions and on their structure. Such study is conducted in the present work by means of the Flory model [20] and the application of the formalism of the Kirkwood-Buff integrals [21-22]. Particularly, the considered esters are of the type $CH_3(CH_2)_uCOO(CH_2)_vCH_3$: methyl alkanoates ($v$ = 0; $u$ = 0-4); ethyl ($v$ = 1), propyl ($v$ = 2) or butyl ($v$ = 3) alkanoates ($u$ = 0-2) and pentyl alkanoates ($v$ = 4; $u$ = 0,1). In addition, proximity effects in systems with diesters are also investigated. This work is a continuation of previous research where the Flory model was applied for the study of mixtures containing the O and/or CO groups: ethers [23], or *n*-alkanones [24], or *n*-alkanals [24], or dialkyl carbonates [24,25] and alkanes. For the sake of comparison and in order to complete our study, the Kirkwood-Buff formalism has also been -applied to some *n*-alkanone or dialkyl carbonate + heptane mixtures. Finally, it should be mentioned that, in the framework of the Flory model, some studies, mainly concerned with the correlation/prediction of excess molar volumes ($V_m^E$), of ester + alkane mixtures are available in the literature [15,26,27].

## 2. Theories

### *2.1 Flory model*

A short summary of the essential hypotheses of the theory [20,28-30] can be encountered elsewhere [31]. The basic assumption of the model is that of random mixing. The Flory equation of state is:

$$\frac{\hat{P}\hat{V}}{\hat{T}} = \frac{\hat{V}^{1/3}}{\hat{V}^{1/3}-1} - \frac{1}{\hat{V}\hat{T}} \quad (1)$$

where $\hat{V} = V_{\mathrm{m}} / V_{\mathrm{m}}^*$; $\hat{P} = P / P^*$ and $\hat{T} = T / T^*$ are the reduced volume, pressure and temperature, respectively ($V_{\mathrm{m}}$ is the molar volume of the mixture). Equation (1) is valid for pure liquids and liquid mixtures. For pure liquids, the reduction parameters, $V_{\mathrm{mi}}^*$, $P_{\mathrm{i}}^*$ and $T_{\mathrm{i}}^*$ can be obtained from experimental data, such as density, $\alpha_{p\mathrm{i}}$ (isobaric expansion coefficient) and $\kappa_{T\mathrm{i}}$ (isothermal compressibility). Expressions for reduction parameters of mixtures have been given previously [31]. $H_{\mathrm{m}}^{\mathrm{E}}$ values are determined from:

$$H_{\mathrm{m}}^{\mathrm{E}} = \frac{x_1 V_{\mathrm{m1}}^* \theta_2 X_{12}}{\hat{V}} + x_1 V_{\mathrm{m1}}^* P_1^* \left(\frac{1}{\hat{V}_{\mathrm{m1}}} - \frac{1}{\hat{V}}\right) + x_2 V_{\mathrm{m2}}^* P_2^* \left(\frac{1}{\hat{V}_{\mathrm{m2}}} - \frac{1}{\hat{V}}\right) \quad (2)$$

where all the symbols have their usual meaning [31]. The reduced volume of the mixture, $\hat{V}$, is obtained from the equation of state. Therefore, the molar excess volume can be also calculated:

$$V_{\mathrm{m}}^{\mathrm{E}} = (x_1 V_{\mathrm{m1}}^* + x_2 V_{\mathrm{m2}}^*)(\hat{V} - \varphi_1 \hat{V}_{\mathrm{m1}} - \varphi_2 \hat{V}_{\mathrm{m2}}) \quad (3)$$

### 2.2 Kirkwood-Buff integrals

In this formalism, the Kirkwood-Buff integrals are determined from [32,33]:

$$G_{\mathrm{ij}} = \int_o^\infty (g_{\mathrm{ij}} - 1) 4\pi r^2 dr \quad (4)$$

where, $g_{\mathrm{ij}}$ is the radial distribution function (probability of finding a molecule of species i in a volume element at the distance $r$ of the center of a molecule of type j). Thus, $g_{\mathrm{ij}}$ provides information about the mixture structure at microscopic level. Values of $G_{\mathrm{ij}}$ can be interpreted as follows: positive values represent the excess of molecules of the type i in the space around a given molecule of kind j. That is, attractive interactions between molecules of i and j exist. Negative values of $G_{\mathrm{ij}}$ reveal that i-i and j-j interactions are predominant over i-j interactions [32,34]. The Kirkwood-Buff integrals can be derived from thermodynamic data such as chemical potential; partial molar volumes and isothermal compressibility factor. The resulting equations are [32,35]:

$$G_{\mathrm{ii}} = RT\kappa_T + \frac{x_{\mathrm{j}} \overline{V}_{\mathrm{mj}}^2}{x_{\mathrm{i}} V_{\mathrm{m}} D} - \frac{V_{\mathrm{m}}}{x_{\mathrm{i}}} \text{ (i,j=1,2)} \quad (5)$$

$$G_{12} = G_{21} = RT\kappa_T - \frac{\bar{V}_{m1}\bar{V}_{m2}}{V_m D} \quad (6)$$

where *R* is the gas constant, $x_i$ and $\bar{V}_{mi}$ are the mole fraction and the partial molar volume of component i, respectively (i = 1,2) and $\kappa_T$, the isothermal compressibility of the mixture. *D* is defined as:

$$D = 1 + \frac{x_1 x_2}{RT}\left(\frac{\partial^2 G_m^E}{\partial x_1^2}\right)_{P,T} \quad (7)$$

In this expression, $G_m^E$ denotes the excess molar Gibbs energy. The $G_{ij}$ integrals allow estimate the linear coefficients of preferential solvation [36]:

$$\delta_{ii}^0 = x_i x_j (G_{ii} - G_{ij}) \quad (8)$$
$$\delta_{ij}^0 = x_i x_j (G_{ij} - G_{jj})$$

which are useful quantities to determine the local mole fractions of the i species around the central j molecule [36].

## 3 Model calculations and results

Values of physical properties of *n*-alkanoates, required for calculations, are collected in Table 1. For other organic solvents considered, *n*-alkanes, *n*-alkanones and linear organic carbonates the values used have been taken from previous applications [25]. In order to calculate $P_i^*, V_i^*, T_i^*$ (i = A,B) at $T \neq 298.15$ K, expressions provided in reference [37] for the temperature dependence of density, $\alpha_p$ and $\gamma$ (= $\alpha_p/\kappa_T$) were used.

### *3.1 Flory*

Table 2 lists values of the interaction parameter $X_{12}$ determined from the corresponding experimental $H_m^E$ results at equimolar composition and 298.15 K following the method provided in reference [38]. Table 2 also includes the relative standard deviations for $H_m^E$ defined as:

$$\sigma_r(H_m^E) = \left[\frac{1}{N}\sum\left(\frac{H_{m,exp}^E - H_{m,calc}^E}{H_{m,exp}^E}\right)^2\right]^{1/2} \quad (9)$$

where $N$ is the number of experimental data points (see Figures 1-2). Figure S1 (supplementary material) shows the concentration dependence of $X_{12}$ for a few systems. Such dependence has been determined according to the method explained in detail elsewhere [39]. Results provided by the model on $V_{\rm m}^{\rm E}$ are collected in Table S1.

*3.2 Kirkwood-Buff integrals*

In these calculations, isothermal compressibilities of the mixtures were determined from $\kappa_T = \Phi_1\kappa_{T1} + \Phi_2\kappa_{T2}$ (where $\Phi_i = x_{\rm i}V_{\rm mi}/(x_1V_{\rm m1} + x_2V_{\rm m2})$ is the volume fraction of the component i of the system), i.e., ideal behaviour is assumed). This is a typical approach, which has not influence on the final calculations of the Kirkwood-Buff integrals [40]. Along calculations, $G_{\rm m}^{\rm E}$ values were determined from the DISQUAC model using interaction parameters from the literature [18,41,42]. Results on $\delta_{\rm ij}^0 = x_{\rm i}x_{\rm j}(G_{\rm ij} - G_{\rm jj})$ are collected in Table 3 (Figures 3a,3b), together with the source of experimental data on $V_{\rm m}^{\rm E}$ used.

## 4. Discussion

*4.1 $CH_3(CH_2)_u$ $COO(CH_2)_vCH_3$ (1) + n-alkane(2)*

*4.1.1. Systems with heptane*

$H_{\rm m}^{\rm E}$ results of this type of mixtures are positive indicating that interactions between like molecules are dominant. On the other hand, $H_{\rm m}^{\rm E}$ values decrease along a given homologous series ($u$ = constant when $v$ increases, e.g.). This can be explained assuming that: (i) interactions between ester molecules become weaker, as it is suggested by the corresponding partial excess molar enthalpies at infinite dilution of the *n*-alkanoate, $H_{\rm m1}^{\rm E,\infty}$, which also decrease along each considered series (Table 4); (ii) increasing steric effects, since the COO group becomes more sterically hindered when the ester size increases, and this leads to a decrease of the number of ester-ester interactions in the system in comparison to that of pure *n*-alkanoates [43]. Such effects are the cause of the decrease of the enthalpic parameters when the quasichemical approximation of DISQUAC was applied to the systems under study [19]. In addition, it has been suggested that the decrease of $H_{\rm m}^{\rm E}$ might also be due to long *n*-alkanoates can form quasi-cycles [43]. We note that ester-ester interactions are stronger for the mixture including methyl ethanoate ($u$ = 0, $v$ = 0), characterized by the highest $H_{\rm m1}^{\rm E,\infty}$ value (8.3 kJ mol$^{-1}$) [44]. In fact, systems formed by this ester and *n*-alkane show LLE curves with relatively high upper critical solution temperatures (UCST), 241.7 K for the mixture with octane [45]. Solid-liquid equilibria measurements also reveal partial miscibility for the ethyl ethanoate + *n*-alkane systems at lower temperatures [19]. For example, the UCST for the heptane solution is $\approx 200$ K [19]. The weakening of the ester-ester interactions may be due to the weakening of the dipolar interactions involved. The effective dipole moment $\bar{\mu}_{\mathrm{i}}$ is a useful magnitude to evaluate the impact of polarity on bulk properties [46-48]. For a compound with dipole moment in gas phase $\mu_{\mathrm{i}}$, $\bar{\mu}_{\mathrm{i}}$ is defined according to [48]:

$$\bar{\mu}_{\mathrm{i}} = [\frac{\mu^2 N_{\mathrm{A}}}{4\pi\varepsilon_{\mathrm{o}} V_{\mathrm{mi}} k_{\mathrm{B}} T}]^{1/2} \tag{10}$$

where $N_{\mathrm{A}}$ is the Avogadro's number, $\varepsilon_{\mathrm{o}}$ the permittivity of the vacuum and $k_{\mathrm{B}}$ the Boltzmann's constant. While for a given series, here linear alkanoates, $\mu_{\mathrm{i}}$ varies only slightly with the chain length (Table S2), $\bar{\mu}_{\mathrm{i}}$ shows a much greater variation (Table S2). For example, for $u$ = 0 (alkyl ethanoates) or $v$ = 0 (methyl alkanoates), $\bar{\mu}_{\mathrm{i}}$ decreases when the size of the ester increases, indicating that the dipolar interactions become weaker. The experimental results of the isobaric excess molar heat capacities, $C_{p\mathrm{m}}^{\mathrm{E}}$, for methyl alkanoate + heptane systems support this statement. Thus, the solution with methyl ethanoate shows a W-shaped $C_{p\mathrm{m}}^{\mathrm{E}}$ curve with positive values at the central part of the concentration range (0.91 J mol$^{-1}$ K$^{-1}$) [49]. This is a typical feature of mixtures that, at the working temperature, 298.15 K, are not far from the UCST [50]. The same trend is observed for the mixtures propanone + dodecane [51], or dimethyl carbonate (DMC) + octane, or + nonane [52]. For systems with longer methyl $n$-alkanoates, the ($C_{p\mathrm{m}}^{\mathrm{E}}$/J mol$^{-1}$ K$^{-1}$) results become more negative when the ester size increases: −0.29 (methyl propanoate); −0.75 (methyl octanoate) [49], and, similarly, −0.50 for the mixture with propyl butanoate [53]. Negative $C_{p\mathrm{m}}^{\mathrm{E}}$ values are encountered in systems where dispersive interactions are dominant, e.g., benzene + $n$-alkane [54,55]. Inspection of Table 4 reveals that for a number of mixtures, the $H_{\mathrm{m1}}^{\mathrm{E},\infty}$ values are quite similar (propyl, butyl propanoate or butanaote) and the same occurs for $\bar{\mu}_{\mathrm{i}}$, while the $H_{\mathrm{m}}^{\mathrm{E}}$ results are different. For example, $H_{\mathrm{m1}}^{\mathrm{E},\infty}$/kJ mol$^{-1}$ $\approx$ 4.4 ($u$ =1,$v$ =2; $u$ =1, $v$ =3); $\bar{\mu}_{\mathrm{i}}$ = 0.591 ($u$ =1,$v$ =2); 0.562 ($u$ =1,$v$ =3) (Table 4) and $H_{\mathrm{m}}^{\mathrm{E}}$/J mol$^{-1}$ = 893 ($u$ =1,$v$ =2) [56]; 758 ($u$ =1,$v$ =3) [57]. It suggests that structural effects may have a relevant impact on the $H_{\mathrm{m}}^{\mathrm{E}}$ variation. In order to investigate this point, properties at constant volume are now considered, particularly the isochoric excess molar internal energy function, $U_{V\mathrm{m}}^{\mathrm{E}}$, which can be obtained from the equation [47]:

$$U_{Vm}^{E} = H_{m}^{E} - T\frac{\alpha_p}{\kappa_T}V_{m}^{E} \quad (11)$$

where $V_{m}^{E}$ is the excess molar volume, and $\alpha_p$ and $\kappa_T$ are, respectively, the isobaric thermal expansion coefficient and the coefficient of isothermal compressibility of the considered system. The $T\frac{\alpha_p}{\kappa_T}V_{m}^{E}$ term describes the contribution from the equation of state (eos) term to $H_{m}^{E}$. If the needed data are not available, $\alpha_p$ and $\kappa_T$ values can be calculated assuming that the mixtures behave ideally with regards to these properties. That is, $M^{id} = \Phi_1 M_1 + \Phi_2 M_2$; with $M_i = \alpha_{pi}$, or $\kappa_{T_i}$. Such approximation is commonly acceptable. It is interesting to observe that $U_{Vm1}^{E,\infty}$ values are nearly constant for mixtures including alkyl propanoates (*u* = 1) with *v* = 1,2,3, or containing ethyl alkanaotes (*v* = 1) with *u* = 1,2 (Table 4). In the former case, $U_{Vm1}^{E,\infty}(u=1) \simeq 3.7$ kJ mol$^{-1}$ for *v* =1,2,3 and $U_{Vm}^{E}$/J mol$^{-1}$ = 896 (*v* = 1), 736 (*v* = 2), 654 (*v* = 3) (Table 4). This variation is probably due to increased steric effects (*v* increases) lead, as already mentioned, to a less number of ester-ester interactions in the system. In contrast, such effects are similar in solutions including an ester characterized by *u* + *v* = 3 (*u* =1, *v* = 2, or *u* = 2, *v* = 1), or by *u* + *v* = 4 (*u* = 4, *v* = 0; *u* = 0, *v* = 4). It is remarkable that, e.g., the COO group is more sterically hindered in the mixture with butyl propanoate (*u* = 1, *v* = 3; $U_{Vm}^{E}$/J mol$^{-1}$ = 654) than in solutions with methyl hexanoate (*u* = 4, *v* = 0) or pentyl ethanoate (*u* = 0, *v* = 4) ($U_{Vm}^{E}$/J mol$^{-1}$ = 729). Note that there is no meaningful difference between the $U_{Vm1}^{E,\infty}$ values of these systems ($\approx$3.7 kJ mol$^{-1}$ (Table 4)). The excess molar volumes $V_{m}^{E}$ are also positive (Table S1) and change in line with the $H_{m}^{E}$ results. Some examples follow, $V_{m}^{E}$/cm$^3$ mol$^{-1}$= 1.384 (methyl ethanoate); 0.295 (methyl hexanoate) [58], and, in the same order, $H_{m}^{E}$/J mol$^{-1}$ = 1787; 838 [14]. It is possible to conclude that the main contribution to $V_{m}^{E}$ arises from interactional effects. The positive values of $((\frac{\Delta V_{m}^{E}}{\Delta T})_p$/cm$^3$mol$^{-1}$K$^{-1}$) agree with such statement: 0.024 (ethyl ethanoate) [59,60], 6.0 10$^{-3}$ (methyl propanoate) [49,61], or 3.0 10$^{-3}$ (propyl propanoate) [56,62].

*4.1.2. Systems with a given n-alkanoate*

In this case, both excess functions $H_{m}^{E}$ and $V_{m}^{E}$ increase with the size of the *n*-alkane (Tables 2 and S1). That is, the variation of $V_{m}^{E}$ can be ascribed to an increase of the corresponding interactional effects. For the mixtures listed in Table S1 (see Figure 4), $U_{Vm}^{E}$ changes similarly. However, some previous calculations show that such variation is not hold for systems with long methyl alkanoates. This matter deserves a careful investigation and will be considered in a forthcoming work.

*4.2. $CH_3CH_2COO(CH_2)_uCOOCH_2CH_3$ + hexane mixtures*

The data needed to conduct a discussion in similar terms that above are not available and only enthalpic data will be here considered. Nevertheless, some quite general trends can be outlined. Firstly, it is to be noted that the inclusion of a second COO group in the aliphatic chain of the *n*-alkanoate leads to higher $H_m^E$ values in alkane solutions. Thus, $H_m^E$(hexane)/J mol$^{-1}$ = 1036 (ethyl propanoate) [63] < 1657 (diethyl oxalate, $T$ = 303.15 K) [64] which can be ascribed to interactions between diester molecules are stronger as the corresponding $H_{m1}^{E,\infty}$ values reveal: $H_{m1}^{E,\infty}$(hexane)/kJ mol$^{-1}$ = 5.7 (ethyl propanoate) [63] < 9.1 (diethyl oxalate) [64]. It is interesting to note that although the $\bar{\mu}_i$ value of diethyl malonate (0.796) is higher than that of butyl propanoate (ester of similar size, 0.565, Table S2), indicating that dipolar interactions between diester molecules are stronger, the $\bar{\mu}_i$ values of other alkanoates are rather similar: 0.555 (diethyl succinate), 0.585 (ethyl hexanoate) (Table S2). This underlines that some additional effect must be considered in order to explain the stronger interactions between diester molecules. Such effect is the so-called proximity effect and is also encountered, e.g., in alkane mixtures with dioxaalkanes [65], or dichloroalkanes [66], or with diketones [67]. In the case of hexane systems, $H_m^E$/J mol$^{-1}$ = 1254 (2-butanone) [68]; 1600 (2,3-butanedione, $T$ = 303.15 K) [67], and 496 (1-chlorobutane) [69], 1245 (1,4-dichlorobutane) [70]. On the other hand, the $H_m^E$ results of $CH_3CH_2COO(CH_2)_uCOOCH_2CH_3$ + hexane mixtures at 303.15 K decrease when $u$ increases ($u \geq 1$): $H_m^E$ /J mol$^{-1}$ = 1707 ($u$ = 1) > 919 ($u$ = 8) [64]. Such variation can be linked to a progressive weakening of the proximity effect, i.e., to weaker ester-ester interactions at the mentioned condition. In fact, the ($H_{m1}^{E,\infty}$ /kJ mol$^{-1}$) values change in the same order [64]: 11.5 ($u$ =1) > 4.8 ($u$ = 8). Mixtures with hexane and dichloroalkanes behave similarly: 1616 (dichloroethane) [71]; 990 (1,6-dichlorohexane) [70]. Nevertheless, the system with $u$ = 0, shows a $H_m^E$ value (1658 J mol$^{-1}$) slightly lower than that corresponding to the mixture with $u$ = 1. It is to be mentioned that diketone + hexane mixtures at 303.15 K behave differently: $H_m^E$ values decrease when replacing 2,3-butanedione by 2,4-pentadione but the system with 2,5-hexanedione shows a miscibility gap at the mentioned temperature [67]. This has been explained in terms of the percentage of enol form of 2,5-hexanedione in hexane is higher than for the other two diketones [67].

*4.3 Comparison with other mixtures*

For solutions with a given *n*-alkane, e.g. decane, UCST/K changes in the order: 255 (methyl ethanoate, estimated from data from [45]) < 266.8 (propanone) [72] < 286.6 (DMC) [73]. It is remarkable that the $\bar{\mu}_i$ value of DMC (0.391, [24]) is lower than those of acetone (1.281 [24]), or of methyl ethanoate (0.737 (Table S2). This suggests that non-random mixing is not only determined by polarity [74] and that other effects (polarizability, size of the polar group) may also contribute to non-random effects. It is known that internal pressure ($P_{\text{int}} = (T\alpha_p / \kappa_T) - p$) is a magnitude mainly determined by dispersive interactions and weak dipole-dipole interactions, and that $P_{\text{int}}V_{\text{m}}$ is a measure of the London dispersive energies [25,75]. Here we have, $P_{\text{int}}V_{\text{m}}$/kJ mol$^{-1}$ = 35.46 (DMC), 29.53 (methyl ethanoate); 24.27 (acetone). That is, dispersive interactions are meaningfully more relevant in DMC, the compound characterized by a larger dipolar polarizability: 7.7 $10^{-24}$ (DMC); 6.9 $10^{-24}$ (methyl ethanoate); 6.6 $10^{-24}$ (acetone) (all values in cm$^3$) [76]. The values of $H_{\text{m1}}^{\text{E},\infty}$ and $U_{V\text{m1}}^{\text{E},\infty}$ for the mixtures with heptane change similarly to the UCST results given above: $H_{\text{m1}}^{\text{E},\infty}$ /kJ mol$^{-1}$ = 8.3 (methyl methanoate) [44] < 9.1 (propanone) [77] < 9.5 (DMC) [78] and $U_{V\text{m1}}^{\text{E},\infty}$/kJ mol$^{-1}$ = 5.7 (methyl methanoate) < 7.1 (propanone) < 8.1 (DMC) (Tables 4,5). One can conclude that interactions between like polar molecules become weaker in the sequence: DMC > propanone > methyl ethanoate. The variation of $H_{\text{m}}^{\text{E}}$ and $U_{V\text{m}}^{\text{E}}$ is different. Thus, $H_{\text{m}}^{\text{E}}$ (heptane)/J mol$^{-1}$ = 1988 (DMC) [78] > 1787 (methyl ethanoate) [14] > 1676 (propanone) [77] and $U_{V\text{m}}^{\text{E}}$ (heptane)/J mol$^{-1}$ = 1632 (DMC) > 1390 (propanone) > 1362 (methyl ethanoate). The two latter values together with the lower $U_{V\text{m1}}^{\text{E},\infty}$ result for the system with methyl ethanoate suggest that a larger neat number of interactions between like molecules are broken upon mixing in this solution than in the mixture with acetone. Similar trends are observed for systems with diethyl carbonate, 3-pentanone or ethyl propanoate (Tables 4,5). Finally, it should be remarked that for systems with a given *n*-alkane, the differences $\left|U_{V\text{m}}^{\text{E}}(\text{CH}_3\text{COO(CH}_2)_u\text{CH}_3) - U_{V\text{m}}^{\text{E}}(\text{CH}_3\text{CO(CH}_2)_u\text{CH}_3)\right|$ are lower than that between the corresponding $H_{\text{m}}^{\text{E}}$ data (Figure 4). That is, the eos contribution to $H_{\text{m}}^{\text{E}}$ is higher in systems with *n*-alkanoates.

*4.4 Results from the application of models*

*4.4.1 Flory*

Firstly, we note that the $X_{12}$ parameter is positive for all the investigated mixtures (Table 2), indicating that interactions between like molecules are prevalent. On the other hand, $X_{12}$ and $H_{\text{m}}^{\text{E}}$ change in line. For example, in systems with heptane and alkyl ethanoate, both $H_{\mathrm{m}}^{\mathrm{E}}$ and $X_{12}$ decrease with the length of the alkyl term (Table 2) and this may ascribed to the weakening of interactions between ester molecules. In contrast, for butyl ethanoate or methyl pentanoate + heptane mixtures, $H_{\mathrm{m}}^{\mathrm{E}}$ practically remains unchanged, and the same occurs for the corresponding interaction parameters. From the $\sigma_{\mathrm{r}}(H_{\mathrm{m}}^{\mathrm{E}})$ values listed in Table 2, we have determined the mean value, $\bar{\sigma}_{\mathrm{r}}(H_{\mathrm{m}}^{\mathrm{E}})$, for the systems involving *n*-alkanoates ($T$ = 298.15 K), according to the equation:

$$\bar{\sigma}_{\mathrm{r}}(H_{\mathrm{m}}^{\mathrm{E}}) = (\sum \sigma_{\mathrm{r}}(H_{\mathrm{m}}^{\mathrm{E}}))/N_{\mathrm{S}} \qquad (12)$$

where $N_{\mathrm{S}}$ is the number of systems ($N_{\mathrm{S}}$ = 58). The result is $\bar{\sigma}_{\mathrm{r}}(H_{\mathrm{m}}^{\mathrm{E}})$ = 0.070 and this points out that orientational effects are rather weak in this type of solutions. In terms of the model, it is quite difficult to make distinction between mixtures which differ by the monoester, except, perhaps, in the case of systems including methyl ethanoate, where orientational effects seem to be stronger. Thus, $\bar{\sigma}_{\mathrm{r}}(H_{\mathrm{m}}^{\mathrm{E}})$ = 0.095 (methyl ethanoate; $N_{\mathrm{S}}$ = 5); 0.057 (ethyl ethanoate; $N_{\mathrm{S}}$ = 4); 0.057 (propyl ethanoate; $N_{\mathrm{S}}$ = 4); 0.052 (butyl ethanoate; $N_{\mathrm{S}}$ = 4); 0.091 (pentyl ethanoate; $N_{\mathrm{S}}$ = 2). The latter result may be consequence of experimental inaccuracies. Systems including diesters ($T$ = 303.15 K), behave similarly to those with monoesters since $\bar{\sigma}_{\mathrm{r}}(H_{\mathrm{m}}^{\mathrm{E}})$ = 0.078 ($N_{\mathrm{S}}$ = 6). At 318.15 K, for mixtures with monoesters, $\bar{\sigma}_{\mathrm{r}}(H_{\mathrm{m}}^{\mathrm{E}})$ = 0.069 ($N_{\mathrm{S}}$ = 12). That is, in such narrow temperature range, orientational effects remain, in practice, unchanged. It is interesting to consider the position of the investigated mixtures within the $G_{\mathrm{m}}^{\mathrm{E}}$ (excess molar Gibbs energy) vs. $H_{\mathrm{m}}^{\mathrm{E}}$ diagram [79,80]. It is well known that systems characterized by dispersive interactions are situated between the lines $G_{\mathrm{m}}^{\mathrm{E}}$ = 1/3 $H_{\mathrm{m}}^{\mathrm{E}}$ and $G_{\mathrm{m}}^{\mathrm{E}}$ = 1/2 $H_{\mathrm{m}}^{\mathrm{E}}$, while those where dipolar interactions are prevalent are encountered between the lines $G_{\mathrm{m}}^{\mathrm{E}}$ =1/2 $H_{\mathrm{m}}^{\mathrm{E}}$ and $G_{\mathrm{m}}^{\mathrm{E}}$ = $H_{\mathrm{m}}^{\mathrm{E}}$. Solutions where association effects are dominant are situated above the line $G_{\mathrm{m}}^{\mathrm{E}}$ = $H_{\mathrm{m}}^{\mathrm{E}}$. In the present case, the situation of the mixtures within this diagram is slightly above the line $G_{\mathrm{m}}^{\mathrm{E}}$ =1/2 $H_{\mathrm{m}}^{\mathrm{E}}$ and this remarks the relevance of dispersive interactions. Some examples follow for heptane mixtures at 298.15 K using $G_{\mathrm{m}}^{\mathrm{E}}$ values determined according to the DISQUAC model with interaction parameters from the literature [18] and $H_{\mathrm{m}}^{\mathrm{E}}$ results listed in Table 2. Thus, $G_{\mathrm{m}}^{\mathrm{E}}/H_{\mathrm{m}}^{\mathrm{E}}$ = 969/1787 = 0.542 (methyl ethanoate); 770/1510 = 0.510 (ethyl ethanoate); 554/980 = 0.565 (butyl ethanoate); 639/1127 = 0.567 (ethyl propanoate).

On the other hand, $\bar{\sigma}_r(H_m^E)$ values for systems including linear organic carbonates (DMC, DEC; $N_S$ = 8) or *n*-alkanones (propanone, 2-butanone, 2-pentanone, 2-hexanone or 3-pentanone; $N_S$ = 10) are, respectively, 0.109 [24] and 0.156 [25]. The DISQUAC model has been also used to determine $G_m^E$ values of linear organic carbonate, or *n*-alkanone + heptane mixtures [41,42] in order evaluate $G_m^E/H_m^E$ using enthalpic data listed in Table 4. Thus, $G_m^E/H_m^E$ = 0.663 (propanone), 0.603 (DMC); 0.663 (3-pentanone); 0.593 (DEC). This set of results allow to conclude that orientational effects become weaker in the sequence *n*-alkanone > organic carbonate > *n*-alkanoate, which is confirmed by the $\bar{\sigma}_r(H_m^E)$ values for systems including homomorphic polar molecules shown in Figure 5. Accordingly with these results, the variation of the ratio ( $X_{12}$( $x_1$)/ $X_{12}$( $x_1$ = 0.5)) is sharper for mixtures with *n*-alkanones, DMC or DEC than for solutions containing *n*-alkanoates (Figure S1).

The Flory model has been shortly applied to the mixtures 2,3-butanedione or 2,4-pentadione + hexane ($T$ = 303.15 K) [67] and the results are: $X_{12}$/J cm$^{-3}$ = 78.66 (2,3-butanedione); 37.84 (2,4-pentadione) and, in the same order, $\sigma_r(H_m^E)$ = 0.067, 0.148.

*4.4.2 Kirkwood-Buff integrals*

Firstly, we note that for $CH_3COO(CH_2)_vCH_3$ (1) + heptane (2) mixtures at 298.15 K the $\delta_{11}^0$ values decrease when $v$ is increased. This result reveals that the number of ester-ester interactions decreases in systems with alkyl ethanoates when the ester size increases and supports the previous statement about that the observed $H_m^E$ decrease at this condition is partially due to a lower number of interactions between *n*-alkanoate molecules in the mixture. Similar trends are observed in systems with methyl or ethyl alkanoates (Table 3). The effect linked to the increasing of the *n*-alkane size in systems with a given alkanoate is shown in the mentioned Table by means of the results for the methyl ethanoate (1) + dodecane (2) mixture. The most relevant feature of such results is that the $\delta_{11}^0$ curve is skewed towards higher $x_1$ values, which is consistent with the fact that the critical composition of the LLE curves of methyl acetate + *n*-alkane mixtures becomes also shifted to higher concentrations of the alkanoate when the alkane size increases [19,45].

The comparison with results for systems containing *n*-alkanones or organic carbonates shows a decreasing of $\delta_{11}^0$ in the sequence: organic carbonate > *n*-alkanone > *n*-alkanoate (Figures 3a, 3b). This is consistent with the fact that, at 298.15 K, the mixtures with DMC or propanone are closer to the corresponding UCST than the solution with methyl ethanoate.

Finally, some tentative calculations have been conducted for the system 2,3-butanedione + heptane assuming ideal volume of mixing and using $G_{\mathrm{m}}^{\mathrm{E}}$ values from reference [81]. In this case, at equimolar composition and 298.15 K, $\delta_{11}^{0}$ = 1600 $\mathrm{cm}^3\ \mathrm{mol}^{-1}$, which reveals that interactions between ketone molecules are here much more probable than in the solution with 2-butanone.

## 5. Conclusions

The investigated mixtures are mainly characterized by dispersive interactions and by steric effects. The former is supported by: (i) slightly positive or negative values of $C_{pm}^{\mathrm{E}}$; (2) the situation of the systems within the $G_{\mathrm{m}}^{\mathrm{E}}$ vs. $H_{\mathrm{m}}^{\mathrm{E}}$ diagram; (iii) results from the Flory model. The theory cannot make a meaningful distinction between systems with a given *n*-alkane and which differ by the *n*-alkanoate. Steric effects are determinant to explain the observed differences between $H_{\mathrm{m}}^{\mathrm{E}}$ (or $U_{V\mathrm{m}}^{\mathrm{E}}$) values of solutions formed by heptane and isomeric *n*-alkanoates. Orientational effects are also weak in diester + hexane mixtures. It has been shown that, when isomeric polar compounds are considered, the orientational effects become weaker in the sequence: *n*-alkanone > dialkyl carbonate > *n*-alkanoate. The application of the Kirkwood-Buff formalism reveals that the number of ester-ester interactions decrease in systems with alkyl ethanoates when the alkyl size increases and that preferential solvation between polar molecules decreases in the order dialkyl carbonate > *n*-alkanone > *n*-alkanoate.

**Acknowledgements**

The authors gratefully acknowledge the financial support received from the Ministerio de Ciencia e Innovación, under Project PID2022 – 137104NA – 100.

**TABLE 1**

Physical constants and Flory parameters[a] of pure compounds at $T$ = 298.15 K.

| compound | $V_{mi}$ /cm$^3$ mol$^{-1}$ | $\alpha_{pi}$ /10$^{-3}$ K$^{-1}$ | $\kappa_{Ti}$ /10$^{-12}$ Pa$^{-1}$ | $V^*_{mi}$ /cm$^3$ mol$^{-1}$ | $P^*_i$ / J cm$^{-3}$ |
|---|---|---|---|---|---|
| methyl ethanoate | 79.82[b] | 1.42[b] | 1143[b] | 60.16 | 653.7 |
| ethyl ethanoate | 98.47[c] | 1.41[d] | 1191[c] | 74.26 | 620.7 |
| propyl ethanoate | 115.72[e] | 1.27[f] | 1094[f] | 88.98 | 585,4 |
| butyl ethanoate | 132.60[e] | 1.17[f] | 1029[f] | 103.47 | 556.8 |
| pentyl ethanoate | 149.39[c] | 1.15[c] | 995[c] | 117.9 | 527.0 |
| methyl propanoate | 96.92[b] | 1.31[b] | 1077[b] | 74.10 | 620.3 |
| ethyl propanoate | 115.48[g] | 1.3[g] | 1157[g] | 88.41 | 571.5 |
| propyl propanoate | 132.65[h] | 1.21[h] | 1070[h] | 102.89 | 650.4 |
| butyl propanoate | 149.42[i] | 1.14[i] | 990[j] | 117.13 | 558.7 |
| methyl butanoate | 114.42[b] | 1.22[b] | 1031[b] | 88.63 | 588.1 |
| ethyl butanoate | 133.01[k] | 1.17[k] | 1000[k] | 103.72 | 572.9 |
| methyl pentanoate | 131.30[l] | 1.14[l] | 969[j] | 102.92 | 570.8 |
| ethyl pentanoate | 149.52[k] | 1.08[k] | 1200[k] | 118.30 | 428.6 |
| methyl hexanoate | 148.05[b] | 1.08[b] | 944[b] | 117.13 | 545.0 |
| diethyl oxalate | 136.29[m] | 1.1[m] | 770[m] | 107.50 | 684.6 |
| diethyl malonate | 152.81[c] | 0.937[c] | 707[c] | 123.77 | 602.3 |
| diethyl succinate | 168.25[n] | 1.04[n] | 718[n] | 133.97 | 681.1 |
| diethyl glutarate | 184.81[o] | 0.87[o] | 619[j] | 151.43 | 624.1 |
| diethyl adipate | 201.88[p] | 0.96[p] | 693[p] | 163.71 | 608.5 |
| diethyl pimelate | 218.77[q] | 0.86[q] | 634[j] | 179.57 | 600.2 |
| diethyl sebacate | 269.36[r] | 0.87[r] | 663[j] | 220.71 | 582.7 |

[a]$V_{mi}$ , molar volume; $\alpha_{pi}$ , isobaric expansion coefficient; $\kappa_{Ti}$ , isothermal compressibility, $V^*_{mi}$ , $P^*_i$ , reduction parameters for volume and pressure, respectively; [b][26]; [c][82]; [d][60]; [e][83]; [f][84]; [g][85]; [h] [86,87]; [i][88]; [j]estimated using the Manzini-Creszenci method [89]; [k][90]; [l][91]; [m][92,93]; [n][94]; [o][95,96]; [p][97]; [q][95]; [r][98]

**TABLE 2**

Excess molar enthalpies, $H_m^E$, at equimolar composition and temperature *T*, for *n*-alkanoate (1) + *n*-alkane (2) mixtures. Values of the Flory interaction parameter, $X_{12}$, and of the relative standard deviations for $H_m^E$, $\sigma_r(H_m^E)$ (equation 9), are also included.

| System | $N^b$ | *T*/K | $H_m^E$/J mol$^{-1}$ | $X_{12}$ /J cm$^{-3}$ | $\sigma_r(H_m^E)$ | Ref. |
|---|---|---|---|---|---|---|
| methyl ethanoate + *n*-$C_7$ | 15 | 298.15 | 1787 | 92.17 | 0.047 | 14 |
| | 18 | 318.15 | 1944 | 98.68 | 0.075 | 99 |
| methyl ethanoate + *n*-$C_{10}$ | 20 | 298.15 | 2062 | 100.56 | 0.084 | 100 |
| methyl ethanoate + *n*-$C_{12}$ | 21 | 298.15 | 2144 | 102.11 | 0.081 | 59 |
| methyl ethanoate + *n*-$C_{14}$ | 21 | 298.15 | 2363 | 110.44 | 0.113 | 101 |
| methyl ethanoate + *n*-$C_{16}$ | 20 | 298.15 | 2429 | 111.09 | 0.150 | 15 |
| ethyl ethanoate + *n*-$C_7$ | 18 | 298.15 | 1510 | 66.93 | 0.080 | 102 |
| | 15 | 318.15 | 1523 | 66.48 | 0.080 | 99 |
| ethyl ethanoate + *n*-$C_8$ | 18 | 298.15 | 1609 | 69.62 | 0.026 | 60 |
| | 18 | 318.15 | 1623 | 69.37 | 0.036 | 60 |
| ethyl ethanoate + *n*-$C_{10}$ | 18 | 298.15 | 1691 | 70.56 | 0.044 | 60 |
| | 17 | 318.15 | 1757 | 72.79 | 0.066 | 60 |
| ethyl ethanoate + *n*-$C_{12}$ | 20 | 298.15 | 1727 | 76.30 | 0.080 | 59 |
| propyl ethanoate + *n*-$C_6$ | 17 | 298.15 | 1120 | 45.60 | 0.055 | 43 |
| propyl ethanoate + *n*-$C_7$ | 13 | 298.15 | 1199 | 46.95 | 0.032 | 43 |
| | 16 | 318.15 | 1223 | 47.20 | 0.129 | 99 |
| propyl ethanoate + *n*-$C_8$ | 14 | 298.15 | 1281 | 48.69 | 0.052 | 43 |
| | 16 | 318.15 | 1303 | 48.90 | 0.064 | 103 |
| propyl ethanoate + *n*-$C_{10}$ | 15 | 298.15 | 1416 | 51.49 | 0.068 | 103 |
| | 14 | 318.15 | 1454 | 52.41 | 0.059 | 103 |
| butyl ethanoate + *n*-$C_7$ | 21 | 298.15 | 980 | 34.81 | 0.061 | 104 |
| | 18 | 318.15 | 989 | 34.70 | 0.075 | 99 |
| butyl ethanoate + *n*-$C_{10}$ | 20 | 298.15 | 1174 | 38.24 | 0.060 | 104 |
| butyl ethanoate + *n*-$C_{12}$ | 19 | 298.15 | 1303 | 40.95 | 0.057 | 104 |
| butyl ethanoate + *n*-$C_{14}$ | 19 | 298.15 | 1422 | 43.54 | 0.054 | 104 |
| pentyl ethanoate + *n*-$C_7$ | 17 | 298.15 | 828 | 27.20 | 0.086 | 105 |
| | 18 | 318.15 | 881 | 28.59 | 0.098 | 99 |

Table 2 (continued)

| | | | | | | |
|---|---|---|---|---|---|---|
| pentyl ethanoate + *n*-$C_9$ | 18 | 298.15 | 947 | 28.87 | 0.097 | 105 |
| methyl propanoate + *n*-$C_7$ | 15 | 298.15 | 1424 | 63.31 | 0.059 | 14 |
| | 19 | 318.15 | 1514 | 65.89 | 0.042 | 62 |
| methyl propanoate + *n*-$C_8$ | 20 | 298.15 | 1539 | 66.64 | 0.094 | 106 |
| methyl propanoate + *n*-$C_{10}$ | 20 | 298.15 | 1689 | 70.27 | 0.072 | 100 |
| methyl propanoate + *n*-$C_{12}$ | 20 | 298.15 | 1837 | 74.28 | 0.069 | 107 |
| methyl propanoate + *n*-$C_{14}$ | 19 | 298.15 | 1940 | 76.86 | 0.088 | 101 |
| methyl propanoate + *n*-$C_{16}$ | 20 | 298.15 | 2027 | 78.36 | 0.078 | 15 |
| ethyl propanoate + *n*-$C_6$ | 21 | 298.15 | 1036 | 42.28 | 0.047 | 63 |
| ethyl propanoate + *n*-$C_7$ | 16 | 298.15 | 1127 | 44.29 | 0.065 | 102 |
| | 17 | 318.15 | 1153 | 44.66 | 0.048 | 62 |
| ethyl propanoate + *n*-$C_{10}$ | 20 | 298.15 | 1307 | 47.86 | 0.072 | 108 |
| ethyl propanoate + *n*-$C_{12}$ | 20 | 298.15 | 1430 | 50.84 | 0.069 | 108 |
| ethyl propanoate + *n*-$C_{14}$ | 22 | 298.15 | 1541 | 53.64 | 0.056 | 63 |
| propyl propanoate + *n*-$C_6$ | 18 | 298.15 | 821 | 30.57 | 0.099 | 56 |
| propyl propanoate + *n*-$C_7$ | 16 | 298.15 | 893 | 31.76 | 0.091 | 56 |
| | 17 | 318.15 | 907 | 31.84 | 0.061 | 62 |
| propyl propanoate + *n*-$C_8$ | 17 | 298.15 | 944 | 32.46 | 0.076 | 56 |
| propyl propanoate + *n*-$C_{10}$ | 18 | 298.15 | 1068 | 34.98 | 0.056 | 56 |
| butyl propanoate + *n*-$C_7$ | 18 | 298.15 | 758 | 24.89 | 0.100 | 57 |
| butyl propanoate + *n*-$C_9$ | 16 | 298.15 | 871 | 26.66 | 0.073 | 57 |
| methyl butanoate+ *n*-$C_7$ | 17 | 298.15 | 1169 | 46.00 | 0.064 | 14 |
| methyl butanoate + *n*-$C_8$ | 18 | 298.15 | 1241 | 47.33 | 0.041 | 106 |
| methyl butanoate + *n*-$C_{10}$ | 18 | 298.15 | 1376 | 50.10 | 0.072 | 100 |
| methyl butanoate + *n*-$C_{12}$ | 19 | 298.15 | 1494 | 52.66 | 0.076 | 107 |
| methyl butanoate + *n*-$C_{14}$ | 18 | 298.15 | 1583 | 54.33 | 0.064 | 101 |
| methyl butanoate + *n*-$C_{16}$ | 20 | 298.15 | 1665 | 55.84 | 0.057 | 15 |
| ethyl butanoate + *n*-$C_7$ | 16 | 298.15 | 931 | 33.01 | 0.051 | 102 |
| ethyl butanoate + *n*-$C_9$ | 17 | 298.15 | 1101 | 36.65 | 0.113 | 102 |
| methyl pentanoate + *n*-$C_7$ | 18 | 298.15 | 968 | 34.60 | 0.073 | 14 |
| methyl pentanoate + *n*-$C_8$ | 17 | 298.15 | 1019 | 35.11 | 0.074 | 106 |

Table 2 (continued)

| | | | | | | |
|---|---|---|---|---|---|---|
| methyl pentanoate + *n*-$C_{10}$ | 18 | 298.15 | 1140 | 37.25 | 0.068 | 100 |
| methyl pentanoate + *n*-$C_{12}$ | 20 | 298.15 | 1249 | 39.33 | 0.071 | 107 |
| methyl pentanoate + *n*-$C_{14}$ | 19 | 298.15 | 1329 | 40.77 | 0.061 | 101 |
| methyl pentanoate + *n*-$C_{16}$ | 21 | 298.15 | 1409 | 42.00 | 0.057 | 15 |
| ethyl pentanoate + *n*-$C_{7}$ | 17 | 298.15 | 761 | 24.96 | 0.070 | 102 |
| ethyl pentanoate + *n*-$C_{9}$ | 17 | 298.15 | 895 | 27.23 | 0.062 | 102 |
| methyl hexanoate+ *n*-$C_{7}$ | 18 | 298.15 | 838 | 27.69 | 0.058 | 14 |
| methyl hexanoate + *n*-$C_{8}$ | 18 | 298.15 | 888 | 28.13 | 0.085 | 106 |
| methyl hexanoate + *n*-$C_{10}$ | 20 | 298.15 | 983 | 29.33 | 0.080 | 100 |
| methyl hexanoate + *n*-$C_{12}$ | 20 | 298.15 | 1083 | 31.00 | 0.075 | 107 |
| methyl hexanoate + *n*-$C_{14}$ | 20 | 298.15 | 1166 | 32.40 | 0.081 | 101 |
| methyl hexanoate + *n*-$C_{16}$ | 16 | 298.15 | 1233 | 33.24 | 0.064 | 15 |
| diethyl oxalate + *n*-$C_{6}$ | 20 | 303.15 | 1657 | 60.29 | 0.093 | 64 |
| diethyl malonate + *n*-$C_{6}$ | 10 | 303.15 | 1693 | 57.64 | 0.100 | 64 |
| diethyl succinate + *n*-$C_{6}$ | 10 | 303.15 | 1614 | 51.54 | 0.068 | 64 |
| diethyl glutarate + *n*-$C_{6}$ | 10 | 303.15 | 1405 | 43.03 | 0.089 | 64 |
| diethyl adipate + *n*-$C_{6}$ | 12 | 303.15 | 1298 | 37.53 | 0.084 | 64 |
| diethyl sebacate + *n*-$C_{6}$ | 12 | 303.15 | 919 | 23.23 | 0.033 | 64 |

**TABLE 3**

Linear coefficients of preferential solvation, $\delta_{ij}^{0}$, at 298.15 K and composition $x_1$ for *n*-alkanoate(1), *n*-alkanone(1), or linear organic carbonate (1) + heptane (2) mixtures.

| System | $x_1$ | $\delta_{11}^{0}$ /cm$^3$ mol$^{-1}$ | $\delta_{12}^{0}$ /cm$^3$ mol$^{-1}$ | Ref. |
|---|---|---|---|---|
| methyl ethanoate (1) + $n$-C$_7$ (2) | 0.2 | 85.0 | −0.4 | 59 |
| | 0.4 | 181.5 | −45.1 | |
| | 0.6 | 134.7 | −160.1 | |
| | 0.8 | 74.9 | −122.9 | |
| methyl ethanoate (1) + $n$-C$_{12}$ (2) | 0.2 | 78.7 | 18.0 | 59 |
| | 0.4 | 163.8 | 5.0 | |
| | 0.6 | 309.9 | −106.5 | |
| | 0.8 | 463.1 | −573.0 | |
| methyl hexanoate (1) + $n$-C$_7$ (2) | 0.2 | 38.1 | −10.0 | 58 |
| | 0.4 | 45.6 | −31.1 | |
| | 0.6 | 28.3 | −43.1 | |
| | 0.8 | 7.8 | −31.6 | |
| ethyl ethanoate (1) + $n$-C$_7$ (2) | 0.2 | 71.1 | −3.6 | 59 |
| | 0.4 | 121.6 | −38.2 | |
| | 0.6 | 105.9 | −82.0 | |
| | 0.8 | 34.9 | −61.7 | |
| butyl ethanoate (1) + $n$-C$_7$ (2) | 0.2 | 57.5 | −10.6 | 59 |
| | 0.4 | 71.9 | −38.0 | |
| | 0.6 | 44.9 | −52.1 | |
| | 0.8 | 13.2 | −35.6 | |
| ethyl propanoate (1) + $n$-C$_7$ (2) | 0.2 | 61.6 | −6.7 | 102 |
| | 0.4 | 90.3 | −36.3 | |
| | 0.6 | 66.0 | −60.5 | |
| | 0.8 | 21.3 | −43.7 | |
| propanone (1) + $n$-C$_7$ (2) | 0.2 | 164.6 | −9.0 | 77 |
| | 0.4 | 303.9 | −80.4 | |
| | 0.6 | 317.1 | −206.8 | |
| | 0.8 | 80.7 | −122.4 | |

TABLE 3 (continued)

| | | | | |
|---|---|---|---|---|
| 3-pentanone (1) + *n*-$C_7$ (2) | 0.2 | 85.1 | −8.2 | 109 |
| | 0.4 | 113.9 | −40.9 | |
| | 0.6 | 77.5 | −63.2 | |
| | 0.8 | 24.2 | −43.2 | |
| dimethyl carbonate (1) + *n*-$C_7$ (2) | 0.2 | 163.8 | −13.9 | 110 |
| | 0.4 | 515.9 | −181.4 | |
| | 0.6 | 875.2 | −726.1 | |
| | 0.8 | 96.5 | −184.2 | |
| diethyl carbonate (1) + *n*-$C_7$ (2) | 0.2 | 102.7 | −17.2 | 111 |
| | 0.4 | 146.2 | −72.2 | |
| | 0.6 | 88.8 | −96.2 | |
| | 0.8 | 22.8 | −56.2 | |

**TABLE 4**

Excess molar functions at equimolar composition and 298.15 K: enthalpies, $H_{\mathrm{m}}^{\mathrm{E}}$, volumes, $V_{\mathrm{m}}^{\mathrm{E}}$, and isochoric internal energies, $U_{V\mathrm{m}}^{\mathrm{E}}$, and excess molar functions at infinite dilution of the alkanoate at 298.15 K: enthalpies, $H_{\mathrm{m1}}^{\mathrm{E},\infty}$, and internal energies, $U_{V\mathrm{m1}}^{\mathrm{E},\infty}$, for $CH_3(CH_2)_u$ $COO(CH_2)_vCH_3$ (1) + heptane (2) mixtures. The effective dipolar moments, $\bar{\mu}_1$, of *n*-alkanoates are also given (see Table S2).

| | Property | *u* = 0<br>ethanoate | *u* = 1<br>propanoate | *u* = 2<br>butanoate | *u* = 3<br>pentanoate | *u* = 4<br>hexanoate |
|---|---|---|---|---|---|---|
| *v* = 0 | $H_{\mathrm{m}}^{\mathrm{E}}$ / J mol$^{-1}$ | 1787[a] | 1424[b] | 1170[b] | 968[b] | 835[b] |
| methyl | $V_{\mathrm{m}}^{\mathrm{E}}$ /cm$^3$ mol$^{-1}$ | 1.378[c] | 0.973[c] | 0.723[c] | 0.489[c] | 0.295[d] |
| | $U_{V\mathrm{m}}^{\mathrm{E}}$ /J mol$^{-1}$ | 1388 | 1141 | 960 | 826 | 729 |
| | $H_{\mathrm{m1}}^{\mathrm{E},\infty}$ / kJ mol$^{-1}$ | 8.3[a] | 5.4[b] | 5.0[b] | 4.5[b] | 4.2[b] |
| | $U_{V\mathrm{m1}}^{\mathrm{E},\infty}$ / kJ mol$^{-1}$ | 5.7 | 4.3 | 4.0 | 3.8 | 3.5 |
| | $\bar{\mu}_1$ | 0.737 | 0.661 | 0.615 | 0.588 | 0.519 |
| *v* = 1 | $H_{\mathrm{m}}^{\mathrm{E}}$ / J mol$^{-1}$ | 1510[e] | 1127[e] | 931[e] | 761[e] | |
| ethyl | $V_{\mathrm{m}}^{\mathrm{E}}$ /cm$^3$ mol$^{-1}$ | 1.105[f] | 0.803[e] | 0.542[e] | 0.383[e] | |
| | $U_{V\mathrm{m}}^{\mathrm{E}}$ / J mol$^{-1}$ | 1190 | 896 | 772 | 660 | |
| | $H_{\mathrm{m1}}^{\mathrm{E},\infty}$ / kJ mol$^{-1}$ | 7.0[e] | 4.8[e] | 4.5[e] | 3.9[e] | |
| | $U_{V\mathrm{m1}}^{\mathrm{E},\infty}$ / kJ mol$^{-1}$ | 5.1 | 3.7 | 3.7 | 3.1 | |
| | $\bar{\mu}_1$ | 0.700 | 0.591 | 0.581 | | |
| *v* = 2 | $H_{\mathrm{m}}^{\mathrm{E}}$ /J mol$^{-1}$ | 1199[g] | 893[h] | 749[i] | | |
| propyl | $V_{\mathrm{m}}^{\mathrm{E}}$ /cm$^3$ mol$^{-1}$ | 0.787[f] | 0.540[h] | 0.385[i] | | |
| | $U_{V\mathrm{m}}^{\mathrm{E}}$ /J mol$^{-1}$ | 969 | 736 | n.a. | | |
| | $H_{\mathrm{m1}}^{\mathrm{E},\infty}$ / kJ mol$^{-1}$ | 5.6[g] | 4.4[h] | 3.7[i] | | |
| | $U_{V\mathrm{m1}}^{\mathrm{E},\infty}$ / kJ mol$^{-1}$ | 4.6 | 3.7 | | | |
| | $\bar{\mu}_1$ | 0.633 | 0.591 | 0.562 | | |
| *v* = 3 | $H_{\mathrm{m}}^{\mathrm{E}}$ /J mol$^{-1}$ | 980[j] | 758[k] | | | |
| butyl | $V_{\mathrm{m}}^{\mathrm{E}}$ /cm$^3$ mol$^{-1}$ | 0.562[f] | 0.359[k] | | | |
| | $U_{V\mathrm{m}}^{\mathrm{E}}$ /J mol$^{-1}$ | 818 | 654 | | | |
| | $H_{\mathrm{m1}}^{\mathrm{E},\infty}$ / kJ mol$^{-1}$ | 5.4[j] | 4.3[k] | 3.6 | | |
| | $U_{V\mathrm{m1}}^{\mathrm{E},\infty}$ / kJ mol$^{-1}$ | 4.5 | 3.8 | | | |
| | $\bar{\mu}_1$ | 0.622 | 0.565 | 0.563 | | |

Table 4 (continued)

| | | | |
|---|---|---|---|
| $v$ = 4 | $H_{\mathrm{m}}^{\mathrm{E}}$ / J mol$^{-1}$ | 828[l] | 646[l] |
| pentyl | $V_{\mathrm{m}}^{\mathrm{E}}$ /cm$^{3}$ mol$^{-1}$ | 0.347[l] | n.a. |
| | $U_{V\mathrm{m}}^{\mathrm{E}}$ / J mol$^{-1}$ | 729 | |
| | $H_{\mathrm{m1}}^{\mathrm{E},\infty}$ / kJ mol$^{-1}$ | 4.47[l] | 3.4[l] |
| | $U_{V\mathrm{m1}}^{\mathrm{E},\infty}$ / kJ mol$^{-1}$ | 3.7 | n.a. |
| | $\bar{\mu}_1$ | 0.586 | 0.585 |

[a][44]; [b][14]; [c][49]; [d][58]; [e][102]; [f][59]; [g][43]; [h][56]; [i][112]; [j][104]; [k][57]; [l][105]

**TABLE 5**

Excess molar functions at equimolar composition at 298.15 K: enthalpies, $H_{\mathrm{m}}^{\mathrm{E}}$, volumes, $V_{\mathrm{m}}^{\mathrm{E}}$, and isochoric internal energies, $U_{V\mathrm{m}}^{\mathrm{E}}$, and excess molar functions at infinite dilution of the polar compound at 298.15 K: enthalpies, $H_{\mathrm{m1}}^{\mathrm{E},\infty}$, and internal energies, $U_{V\mathrm{m1}}^{\mathrm{E},\infty}$, for *n*-alkanone (1) or linear organic carbonate (2) + heptane (2) mixtures.

| System | $H_{\mathrm{m}}^{\mathrm{E}}$ / J mol$^{-1}$ | $V_{\mathrm{m}}^{\mathrm{E}}$ / cm$^3$ mol$^{-1}$ | $U_{V\mathrm{m}}^{\mathrm{E}}$ / J mol$^{-1}$ | $H_{\mathrm{m1}}^{\mathrm{E},\infty}$ / kJ mol$^{-1}$ | $U_{V\mathrm{m1}}^{\mathrm{E},\infty}$ / kJ mol$^{-1}$ |
|---|---|---|---|---|---|
| propanone (1) + *n*-$C_7$ (2) | 1676[a] | 1.129[a] | 1362 | 9.1[a] | 7.1 |
| 2-butanone (1) + *n*-$C_7$ (2) | 1338[b] | 0.794[c] | 1112 | 7.5[b] | 6.1 |
| 2-pentanone (1) + *n*-$C_7$ (2) | 1135[b] | 0.614[d] | 963 | 6.3[b] | 5.3 |
| 3-pentanone (1) + *n*-$C_7$ (2) | 1078[b] | 0.519[c] | 931 | 5.8[b] | 5.0 |
| 2-hexanone (1) + *n*-$C_7$ (2) | 1055[e] | 0.375[f] | 948 | 6.6[e] | 5.7 |
| DMC (1) + *n*-$C_7$ (2) | 1988[g] | 1.158[h] | 1632 | 9.5[g] | 8.1 |
| DEC (1) + *n*-$C_7$ (2) | 1328[i] | 0.736[j] | 1102 | 7.1[i] | 6.0 |

[a][77]; [b][113]; [c] [114]; [d][115]; [e] [116]; [f][117]; [g][78]; [h][110]; [i] [118]; [j][111]

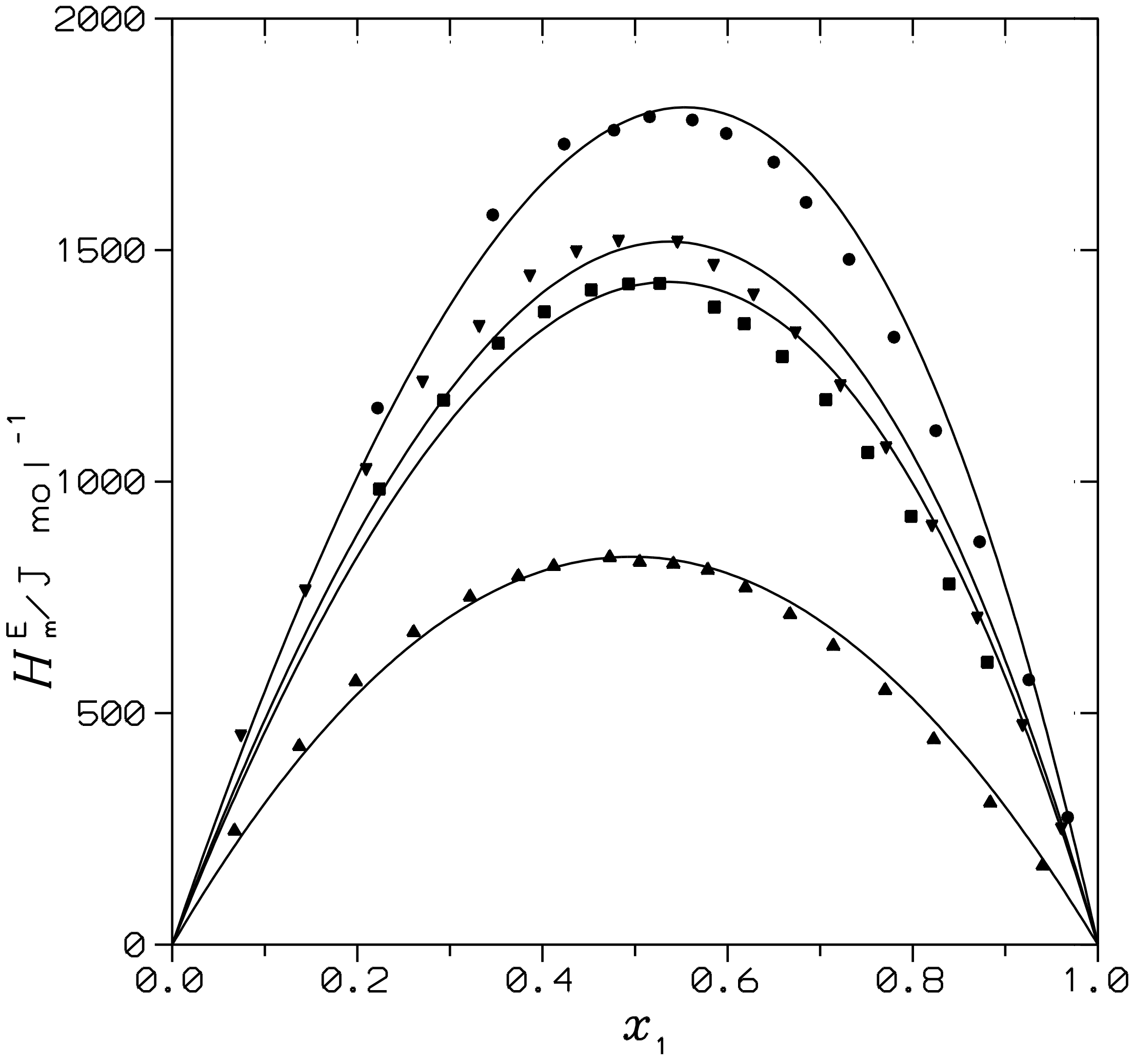

**Figure 1** $H_m^E$ of *n*-alkanoate (1) + heptane (2) systems at 298.15 K. Points, experimental results: (●), methyl ethanoate [14]; (▼), ethyl ethanoate [102]; (■), methyl propanoate [14]; (▲), methyl hexanoate [14]. Solid lines, Flory calculations with interaction parameters listed in Table 2.

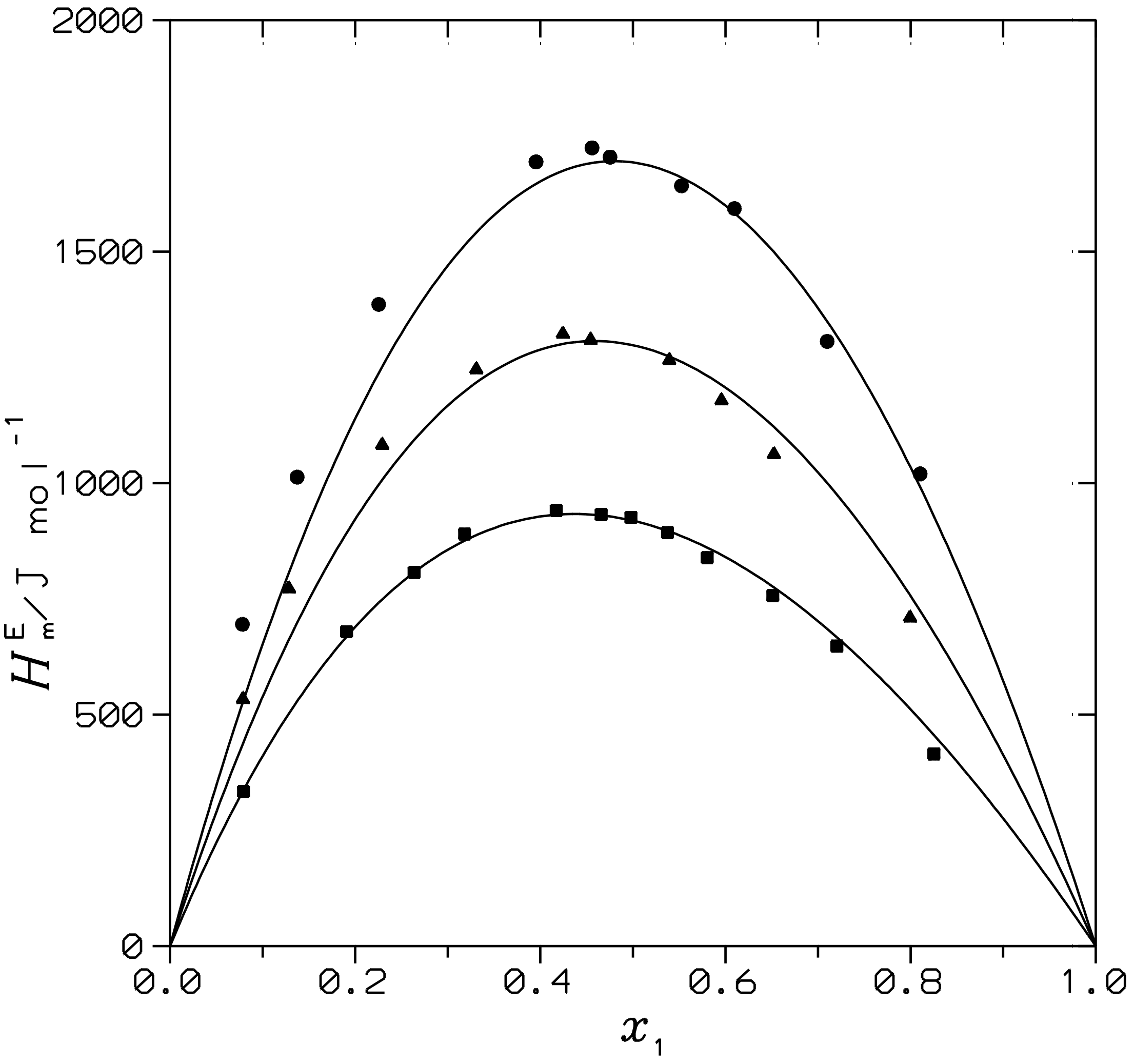

**Figure 2** $H_m^E$ of diester (1) + hexane (2) systems at 303.15 K. Points, experimental results [64]: (●), diethyl malonate; (▲), diethyl adipate; (■), diethyl sebacate. Solid lines, Flory calculations with interaction parameters listed in Table 2.

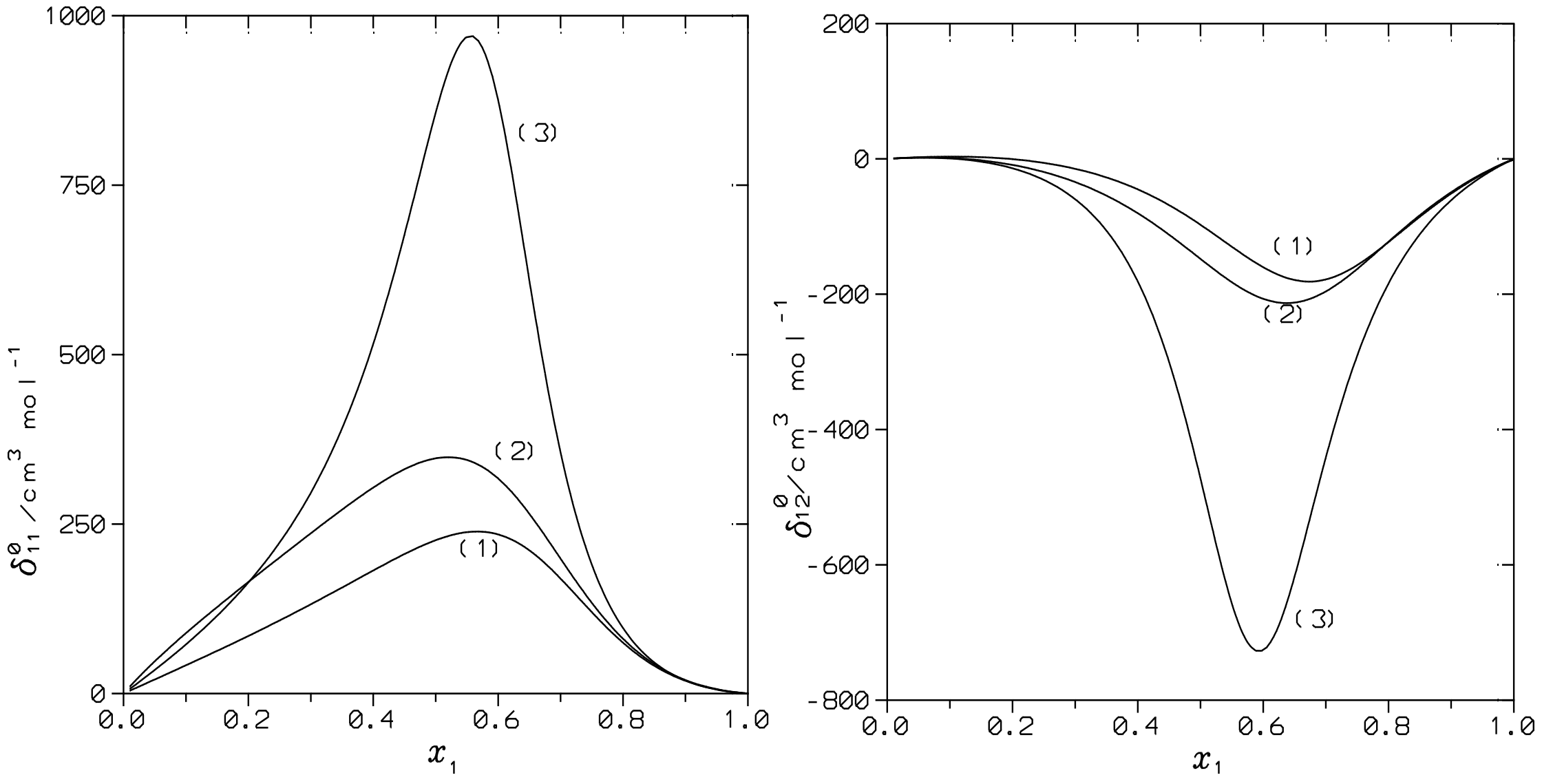

**Figure 3** Linear coefficients of preferential solvation, $\delta^0_{ij}$ , for polar compound (1) + heptane (2) systems at 298.15 K: (1), methyl ethanoate; (2), propanone; (3). dimethyl carbonate. Figure 3a, $\delta^0_{11}$ ; Figure 3b, $\delta^0_{12}$ .

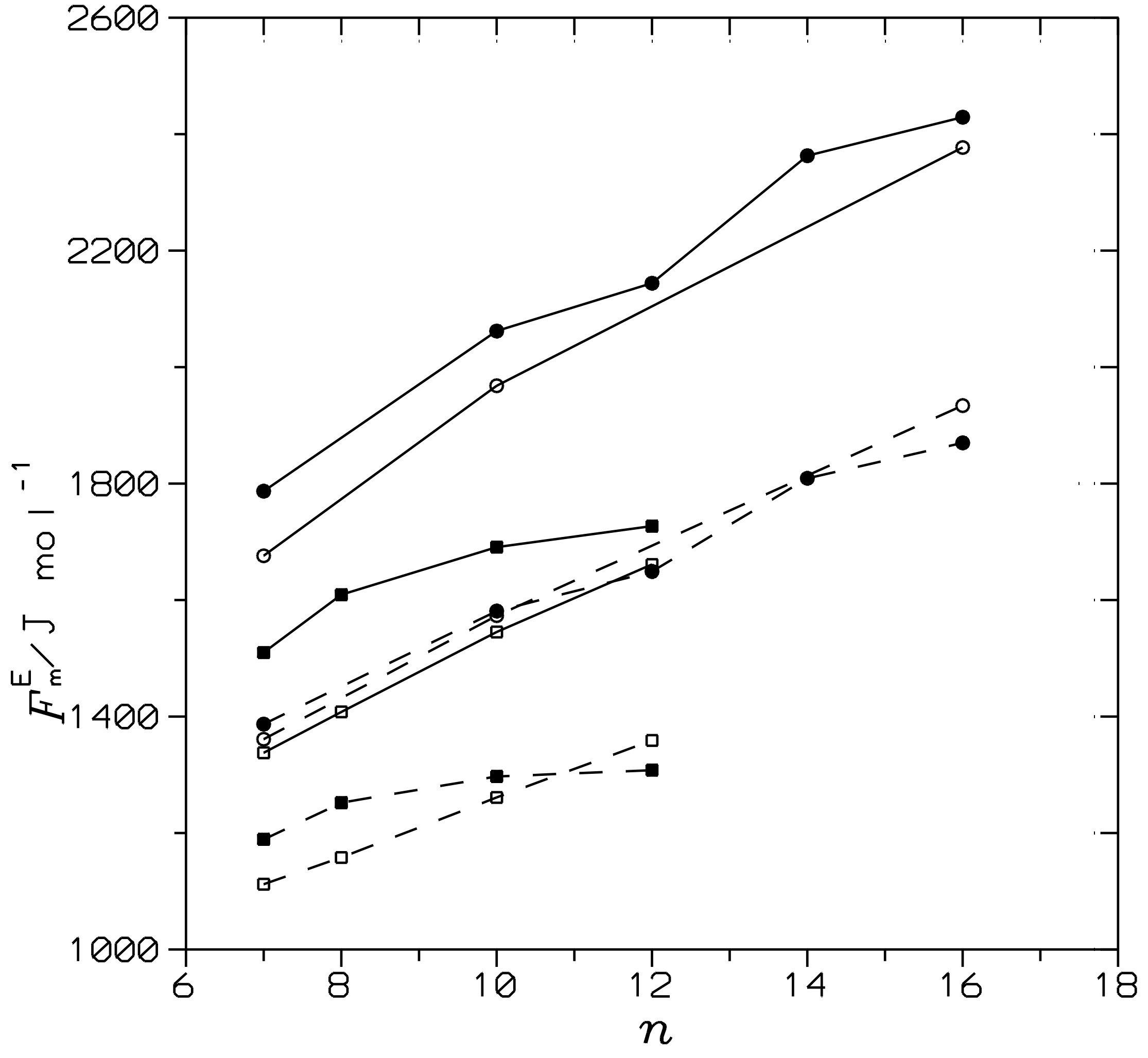

**Figure 4** Excess molar functions, $F_{\mathrm{m}}^{\mathrm{E}}$, at equimolar composition and 298.15 K of linear alkanoate or alkanone + *n*-alkane mixtures vs. *n*, the number of C atoms of the alkane. Lines are for the aid of the eye: solid lines, $H_{\mathrm{m}}^{\mathrm{E}}$; dashed lines, $U_{V\mathrm{m}}^{\mathrm{E}}$. Symbols: (●), methyl ethanoate; (o), propanone; (■), ethyl ethanaote; (□), 2-butanone. For source of experimental $H_{\mathrm{m}}^{\mathrm{E}}$ data, see Tables 2 and 5. Numerical values of $U_{V\mathrm{m}}^{\mathrm{E}}$ are given in Tables S1 and S3.

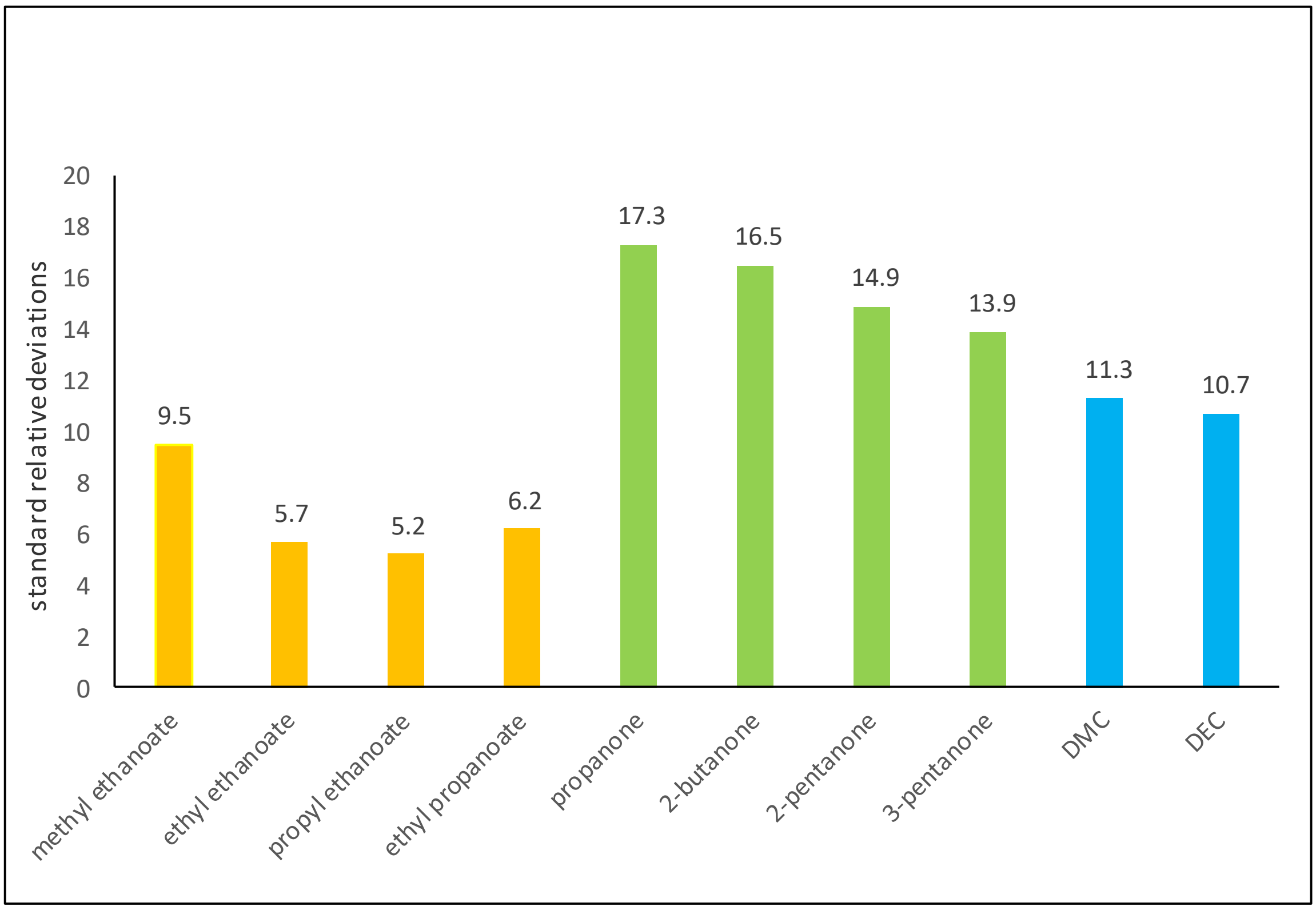

**Figure 5** Standard relative deviations (%) between experimental $H_{\mathrm{m}}^{\mathrm{E}}$ results and values calculated according to the Flory model for polar compound + *n*-alkane systems at 298.15 K: *n*-alkanoates (Table 2); *n*-alkanones [24]; dialkyl carbonates [25].

**SUPPLEMENTARY MATERIAL**

# ORIENTATIONAL AND STERIC EFFECTS IN LINEAR ALKANOATES + *N*-ALKANE MIXTURES

Juan Antonio González,* Fernando Hevia, Luis Felipe Sanz, Daniel Lozano-Martín, Isaías García de la Fuente, José Carlos Cobos

G.E.T.E.F., Departamento de Física Aplicada, Facultad de Ciencias, Universidad de Valladolid, Paseo de Belén, 7, 47011 Valladolid, Spain.

*corresponding author, e-mail: jagl@termo.uva.es; Fax: +34-983-423136; Tel: +34-983-423757

**TABLE S1**

Excess molar volumes, $V_{\mathrm{m}}^{\mathrm{E}}$, and excess molar isochoric internal energies, $U_{V\mathrm{m}}^{\mathrm{E}}$, at 298.15 K and equimolar composition, determined from eq. (11) using $H_{\mathrm{m}}^{\mathrm{E}}$ results from Table 2, for *n*-alkanoate (1) + *n*-alkane (2) mixtures. Results on $V_{\mathrm{m}}^{\mathrm{E}}$ from the Flory model using interaction parameters listed in Table 2 are also included.

| System | *T*/K | $V_{\mathrm{m}}^{\mathrm{E}}$ / cm$^3$mol$^{-1}$ | | $U_{V\mathrm{m}}^{\mathrm{E}}$ / J mol$^{-1}$ | Ref. |
|---|---|---|---|---|---|
| | | Exp. | Flory | | |
| methyl ethanoate + *n*-$C_7$ | 298.15 | 1.378 | 2.047 | 1387 | S1 |
| methyl ethanoate + *n*-$C_8$ | 298.15 | 1.461 | | | S1 |
| methyl ethanoate + *n*-$C_{10}$ | 298.15 | 1.560 | 2.303 | 1581 | S1 |
| methyl ethanoate + *n*-$C_{12}$ | 298.15 | 1.582 | 2.363 | 1649 | S1 |
| methyl ethanoate + *n*-$C_{14}$ | 298.15 | 1.720 | 2.412 | 1809 | S2 |
| methyl ethanoate + *n*-$C_{16}$ | 298.15 | 1.732 | 2.432 | 1870 | S3 |
| ethyl ethanoate + *n*-$C_7$ | 298.15 | 1.105 | 1.741 | 1189 | S1 |
| ethyl ethanoate + *n*-$C_8$ | 298.15 | 1.200 | 1.880 | 1252 | S1 |
| ethyl ethanoate + *n*-$C_{10}$ | 298.15 | 1.283 | 1.938 | 1297 | S1 |
| ethyl ethanoate + *n*-$C_{12}$ | 298.15 | 1.346 | 1.974 | 1308 | S1 |
| propyl ethanoate + *n*-$C_6$ | 298.15 | 0.608 | 0.950 | 955 | S4 |
| propyl ethanoate + *n*-$C_7$ | 298.15 | 0.787 | 1.170 | 971 | S1 |
| propyl ethanoate + *n*-$C_8$ | 298.15 | 0.907 | 1.323 | 1012 | S2 |
| propyl ethanoate + *n*-$C_{10}$ | 298.15 | 1.011 | 1.483 | 1107 | S5 |
| butyl ethanoate + *n*-$C_6$ | 298.15 | 0.335 | | | S6 |
| butyl ethanoate + *n*-$C_7$ | 298.15 | 0.562 | 0.785 | 818 | S1 |
| butyl ethanoate + *n*-$C_8$ | 298.15 | 0.727 | | | S6 |
| butyl ethanoate + *n*-$C_{10}$ | 298.15 | 0.810 | 1.123 | 928 | S6 |
| butyl ethanoate + *n*-$C_{12}$ | 298.15 | 0.899 | 1.276 | 1026 | S6 |
| butyl ethanoate + *n*-$C_{14}$ | 298.15 | 0.982 | 1.318 | 1111 | S6 |
| pentyl ethanoate + *n*-$C_7$ | 298.15 | 0.350 | 0.522 | 726 | S7 |
| ethyl propanoate + *n*-$C_7$ | 298.15 | 0.806 | 1.164 | 895 | S8 |

**TABLE S2**

Dipolar moments, $\mu_{\mathrm{i}}$, [S9] and effective dipolar moments, $\bar{\mu}_{\mathrm{i}}$ (equation 10), at 298.15 K for alkanaoates considered in this work. The corresponding molar volumes, $V_{\mathrm{mi}}$, are the same as in Table 1.

| alkanoate | $V_{\mathrm{mi}}$ /cm$^3$ mol$^{-1}$ | $\mu_{\mathrm{i}}$ /D | $\bar{\mu}_{\mathrm{i}}$ |
|---|---|---|---|
| methyl ethanoate | 79.82 | 1.72 | 0.737 |
| Eehyl ethanoate | 98.47 | 1.82 | 0.702 |
| propyl ethanoate | 115.72 | 1.78 | 0.633 |
| butyl ethanoate | 132.60 | 1.87 | 0.622 |
| pentyl ethanoate | 149.28 | 1.87 | 0.586 |
| methyl propanoate | 96.92 | 1.7 | 0.661 |
| ethyl propanoate | 115.48 | 1.74 | 0.619 |
| propyl propanoate | 132.65 | 1.78 | 0.591 |
| butyl propanoate | 148.59 | 1.8 | 0.565 |
| methyl butanoate | 114.42 | 1.72 | 0.615 |
| ethyl propanoate | 132.97 | 1.75 | 0.581 |
| propyl butanoate | 149.99 | 1.80 | 0.562 |
| methyl pentanoate | 131.30 | 1.76 | 0.588 |
| methyl hexanoate | 148.05 | 1.65 | 0.519 |
| diethyl malonate | 152.81 | 2.57 | 0.796 |
| diethyl succinate | 168.25 | 2.1 | 0.555 |

**Table S3**

Excess molar functions, enthalpies ($H_m^E$), volumes ($V_m^E$) and isochoric internal energies ($U_{Vm}^E$) at 298.15 K and equimolar composition for *n*-alkanone, or dialkyl carbonate + *n*-alkane systems.

| System | $H_m^E$ /J mol$^{-1}$ | $V_m^E$ /cm$^3$ mol$^{-1}$ | $U_{Vm}^E$ /J mol$^{-1}$ |
|---|---|---|---|
| 2-propanone + heptane | 1676 [S10] | 1.130 [S10] | 1361 |
| 2-propanone + decane | 1968 [S11] | 1.333 [S11] | 1573 |
| 2-propanone + hexadecane [a] | 2377 [S11] | 1.422 [S11] | 1934 |
| 2-butanone + heptane | 1338 [S12] | 0.803 [S13] | 1112 |
| 2-butanone + octane | 1408 [S12] | 0.866 [S14] | 1158 |
| 2-butanone + decane | 1545 [S12] | 0.952 [S13] | 1261 |
| 2-butanone + dodecane | 1661 [S15] | 0.996 [S13] | 1359 |
| 2-pentanone + heptane | 1135 [S12] | 0.615 [S16] | 963 |
| 2-pentanone + octane | 1202 [S12] | 0.695 [S17] | 1002 |
| 2-pentanone + decane | 1335 [S12] | 0.812 [S18] | 1094 |
| 3-pentanone + heptane | 1078 [S12] | 0.512 [S19] | 933 |
| 2-hexanone + hexane | 949 [S20] | 0.154 [S21] | 908 |
| 2-hexanone + heptane | 1055 [S20] | 0.375 [S21] | 948 |
| 2-hexanone + octane | 1132 [S20] | 0.516 [S21] | 981 |
| 2-hexanone + nonane | 1198 [S20] | 0.595 [S21] | 1021 |
| 2-hexanone + decane | 1268 [S20] | 0.643 [S21] | 1074 |
| dimethyl carbonate + heptane | 1988 [S22] | 1.158 [S23] | 1642 |
| dimethyl carbonate + decane | 2205 [S22] | 1.442 [S23] | 1747 |
| diethyl carbonate + heptane | 1328 [S24] | 0.736 [S25] | 1101 |
| diethyl carbonate + decane | 1536 [S24] | 1.063 [S25] | 1192 |

[a] There is a partial immiscibility region

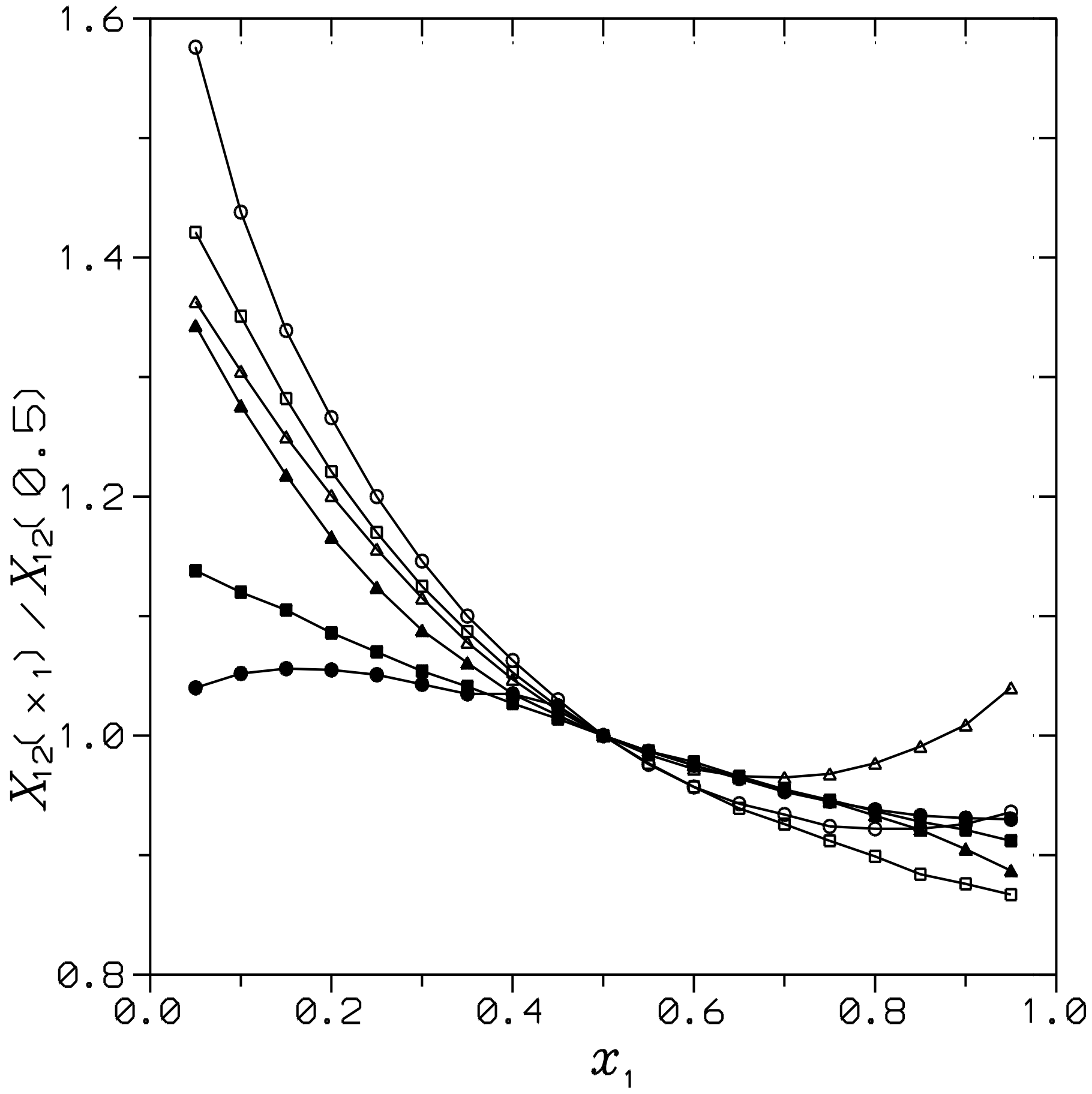

**Figure S1** Ratios of $X_{12}(x_1)$ (interaction Flory parameter at given composition) between $X_{12}(0.5)$ (interaction Flory parameter at equimolar composition, Tables 2 and 5) for polar compound (1) + heptane (2) mixtures at 298.15 K. Symbols: (●), methyl ethanoate; (■), ethyl propanoate; (o), propanone; (□), 3-pentanone; (▲), dimethyl carbonate; (△), diethyl carbonate. Lines are for the aid of the eye.